\documentclass{aa}
\usepackage{graphicx}
\usepackage{txfonts}
\begin{document}
\title{Kinematics and star formation activity in the $z_{\rm abs}$~=~2.03954
damped Lyman-$\alpha$ system towards PKS~0458$-$020\thanks{Based
on observations carried out at the European Southern Observatory (ESO),
under visitor mode progs. ID 66.A-0624, 68.A-0600 and 072.A-0346,
with the UVES echelle spectrograph installed at the ESO Very Large
Telescope (VLT), unit Kueyen, on mount Paranal (Chile).}
}

   \author{Janine Heinm\"uller\inst{1,2,3}
          \and
          Patrick Petitjean\inst{1,4}
          \and
          C\'edric Ledoux\inst{5}
          \and
          Sara Caucci\inst{1}
          \and
          R. Srianand\inst{6}
           }

   \institute{Institut d'Astrophysique de Paris,
              UMR7095 CNRS, Universit\'e Pierre \& Marie Curie,
              98bis Boulevard Arago, 75014, Paris, France
         \and
              Astronomisches Rechen-Institut, Zentrum f\"ur
              Astronomie der Universit\"at Heidelberg, M\"onchhofstr. 12-14,
              69120 Heidelberg, Germany
         \and
             Astrophysik, Universit\"at Potsdam, Am Neuen Palais 10,
              14469 Potsdam, Germany
         \and
              LERMA, Observatoire de Paris, 61 Avenue de l'Observatoire,
              75014, Paris, France
         \and
             European Southern Observatory, Alonso de C\'ordova
             3107, Casilla 19001, Vitacura, Santiago, Chile
         \and
             IUCAA, Post Bag 4, Ganesh Khind, Pune 411 007, India
             }

   \date{Received 25 July 2005; accepted 16 November 2005}

   \abstract{
    We present UVES observations of the log~$N$(H~{\sc i})~=~21.7 damped
    Lyman-$\alpha$ system at $z_{\rm abs}$~=~2.03954 towards
the quasar PKS~0458$-$020.
    H\,{\sc i} Lyman-$\alpha$ emission is detected in the center of
    the damped Lyman-$\alpha$ absorption trough.
    Metallicities are derived for 
Mg~{\sc ii},
Si~{\sc ii},
P~{\sc ii},
Cr~{\sc ii},
Mn~{\sc ii},
Fe~{\sc ii}
and Zn~{\sc ii}
and are found to be
$-1.21\pm0.12$, 
$-1.28\pm0.20$, 
$-1.54\pm0.11$, 
$-1.66\pm0.10$,
$-2.05\pm0.11$,
$-1.87\pm0.11$,
$-1.22\pm0.10$,
respectively, relative to
    solar. The depletion factor is
    therefore of the order of [Zn/Fe$]=0.65$.
    We observe metal absorption lines
    to be blueshifted compared to the Lyman-$\alpha$ emission
   up to a maximum of $\sim$100 and 200~km~s$^{-1}$ for low
   and high-ionization species respectively. This can be
    interpreted either as the consequence of rotation in a
    large ($\sim$7~kpc) disk or as the imprint of a galactic wind.
    The star formation rate (SFR) derived from the
    Lyman-$\alpha$ emission, 1.6~M$_{\odot}$yr$^{-1}$,
    is compared with that estimated from the observed C~{\sc ii}$^*$
    absorption.
   No molecular hydrogen is detected in our data, yielding
   a molecular fraction $\mathrm{log}\,f<-6.52$.
This absence of H$_2$ can be explained as the consequence of a high
ambient UV flux which is one order of magnitude larger than the radiation
field in the ISM of our Galaxy and originates in the observed
emitting region.
   }
   \authorrunning{J. Heinm\"uller et al.}
   \titlerunning{Kinematics and star formation in the DLA towards
PKS~0458$-$020}
   \maketitle
%

\section{Introduction}

 Damped Lyman-$\alpha$ (DLA) systems are characterized by neutral hydrogen
 column densities of $N(\ion{H}{i})~\ge~2~\times~10^{20}$ atoms cm$^{-2}$
determined from the damping wings of the
H~{\sc i} Lyman-$\alpha$
 absorption line. Due to the large \ion{H}{i} column densities
 and the conspicuous presence of metals, DLAs
are believed to arise in intervening galaxies. At low and
 intermediate redshifts, galaxy counterparts have been found
 in a number of cases (Le Brun et al. \cite{lebrun};
 Chen \& Lanzetta \cite{chen}).
Although it is probable that at high redshift DLAs are associated with
regions of star formation, it turns out to be difficult to detect them
in emission. Despite intensive searches, very few cases
 have been found so far in which Lyman-$\alpha$ is seen in emission at
the same redshift as the absorption (e.g., M{\o}ller \&
 Warren \cite{molwar}; M{\o}ller et al. \cite{moller2}; Warren et
 al. \cite{warren}; Vreeswijk et al. \cite{vreeswijk}).
One of these rare cases is the $z_{\rm abs}$~=~2.03954 DLA system towards
PKS~0458$-$020 where
Lyman-$\alpha$ emission from the corresponding absorbing galaxy has
recently been detected by M{\o}ller et al. (\cite{moller})
in the center of the absorption trough.
This DLA system is well known as it was one of the first to be
detected in absorption in 21~cm observations (Wolfe et al. 1985).

 In this paper, we present a new high-resolution spectrum of this quasar
that allows us to discuss the kinematics of
 the Lyman-$\alpha$ emission line relative to the metal lines belonging
 to the DLA system. We measure metallicities and depletion factors.
We discuss and compare two independent methods for the derivation of the
star formation rate, one based on the  Lyman-$\alpha$ emission line and one using
the C~{\sc ii}$^{\rm *}$ absorption line.
 Also, we focus on the physical conditions in the DLA and
 investigate the absence of molecular hydrogen.


\section{Observations and data reduction}

The Ultraviolet and Visible Echelle Spectrograph (UVES; Dekker et al. 2000),
mounted at the Nasmyth B focus of the ESO Kueyen VLT-UT\,2 8.2 m
telescope on
Cerro Paranal in Chile, was used during three visitor mode
observing runs. Dichroic
beam splitters were used on October 21-23, 2000, and October 16, 2001, to
observe at the same time with both spectroscopic arms. During these two
runs, central wavelengths were adjusted to 437 nm in the blue arm and
570, 580
or 750 nm in the red arm. Full wavelength coverage was obtained this
way between 376 and 939 nm with only a small gap between 741 and 757 nm
due to
the physical gap between the two red arm CCDs. The CCD pixels were binned
$2\times 2$ and the slit widths were fixed to $1\arcsec$, yielding, under
the $0\farcs 6$ seeing conditions achieved during the observations, a
resolving
power $R\approx 53,000$. The total integration time on source was about 3.5~h.
Complementary observations at wavelengths
shorter than 387 nm down to the atmospheric cutoff ($\sim 305$ nm)
were obtained on October 29-30, 2003, using the blue arm of UVES in
standalone
together with the standard setting with central wavelength adjusted to
346 nm.
During this third run, due to the faintness of the QSO the CCD pixels were
binned $2\times 3$, while the
slit width again was fixed to $1\arcsec$. These additional observations amount to a total of
about 3.5~h split in three different exposures.

The data were reduced using the latest version of the UVES pipeline
(Ballester et al. 2000) which is available as a dedicated component of the ESO
MIDAS data reduction system. The main characteristics of the pipeline are to
perform a precise inter-order background subtraction for science frames
and master flat-fields, and an optimal extraction with Gaussian modeling of
the object spatial profile rejecting cosmic ray impacts and subtracting
the sky
spectrum simultaneously. The pipeline products were checked step by step.
The wavelength scale of the spectra reduced by the pipeline was then
converted to vacuum-heliocentric values and individual 1-D spectra were
scaled,
weighted and combined to produce the final science spectrum
and its
associated variance.


\section{The Lyman-$\alpha$ emission}\label{lyinem}

\begin{figure}
   \centering
   \includegraphics[bb=44 60 550 730,width=6.5cm,angle=-90,clip]{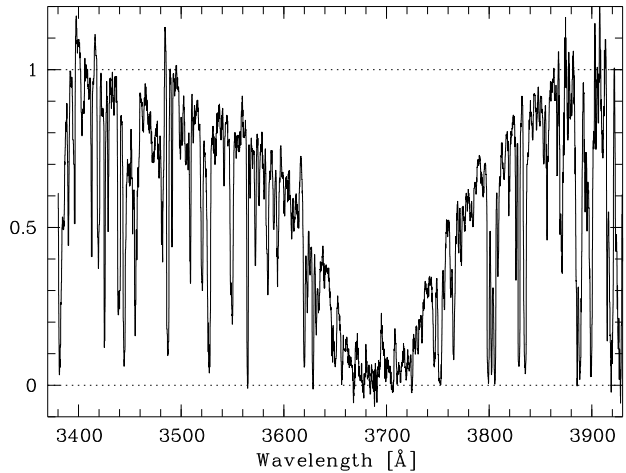}
   \caption{UVES spectrum of PKS~0458$-$020 in the wavelength interval
            between 3400 and 3900 $\rm{\AA}$ showing the damped
            Lyman-$\alpha$ line with the Lyman-$\alpha$ emission line
            in the center. The original spectrum was smoothed with a
            Gaussian filter of FWHM 10 pixels. The neutral hydrogen
            column density of the system, obtained by Voigt profile
            fitting of the absorption trough, is log $N$(H~{\sc
            i})~=~21.7. An emission line is detected in the center of
            the absorption trough.
            }
\label{smoothdla}
    \end{figure}

\begin{figure}
   \centering
   \includegraphics[bb=60 415 320 690,width=6.0cm,clip]{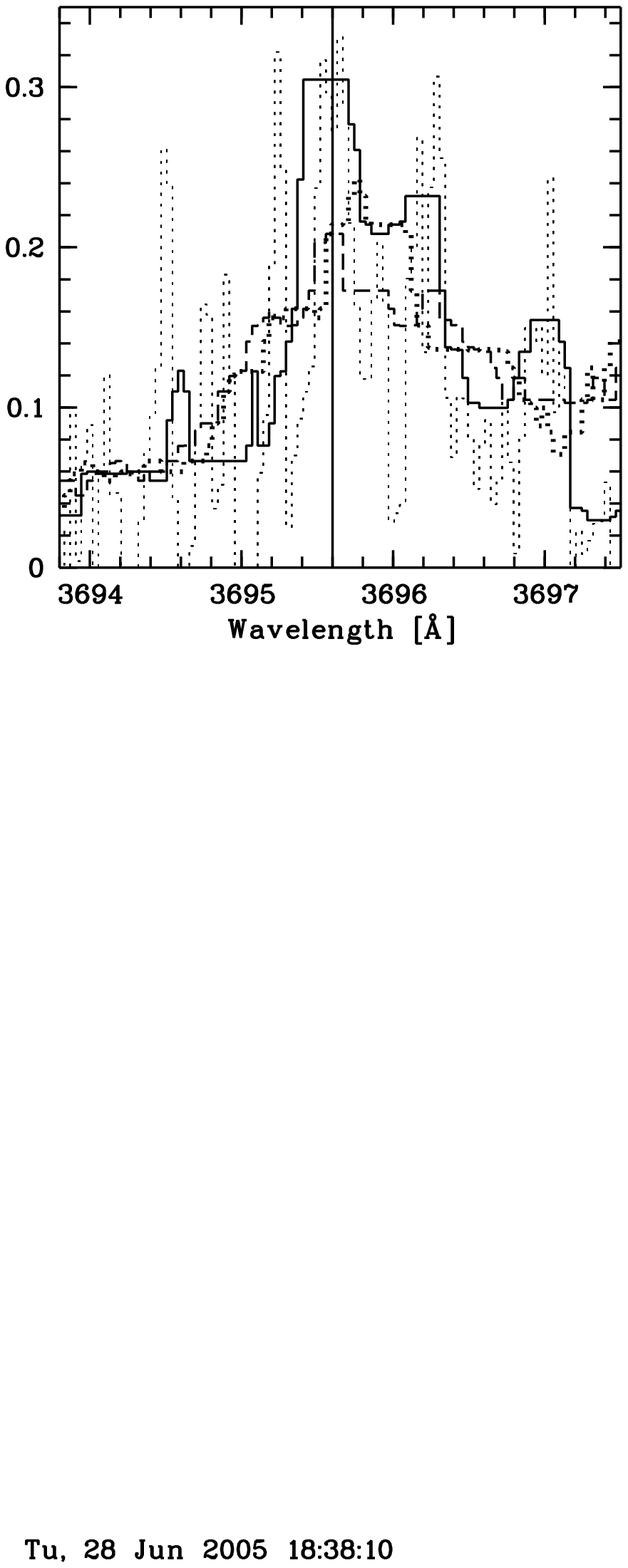}
   \caption{Inset of the spectrum showing the Lyman-$\alpha$ emission
            line centered at 3695.6 $\rm{\AA}$. The thin dotted line is the
            initial high-resolution spectrum. The solid, thick dotted
and dashed lines show the
            emission line smoothed with filter width of 5, 15 and
            20 pixels respectively. Depending on the filter width
            the location of the peak of the emission varies by
$\sim$~0.2 $\rm{\AA}$.
            \label{filter}}
    \end{figure}

By fitting the damped wings, we measure the column density of the damped
Lyman-$\alpha$ absorption to be
log~$N$(\ion{H}{i})~=~21.7$\pm$0.1.
H\,{\sc i} Lyman-$\alpha$ in emission is detected in the
center of the broad absorption as can be seen in Fig. \ref{smoothdla}.
Since the emission line profile is affected by noise we applied
different smoothing factors to the spectrum to measure the exact
wavelength of the Lyman-$\alpha$ emission.
In Fig. \ref{filter}, the initial spectrum of the
Lyman-$\alpha$ emission line is shown together with the results
of applying smoothing with different smoothing radii. As can be
seen the position of the peak of the line
slightly depends on the smoothing radius. The line is not symmetric,
probably because of absorption by intervening neutral hydrogen
in the blue wing. We therefore choose the peak of the line as
an indicator of the mean position of the emission,
3695.6$\pm0.2~\rm{\AA}$ (errors are estimated from the
shifts due to different smoothing). This corresponds to an emission
redshift of $z=2.0400 \pm 0.0002$.
\par\noindent
Our spectrum is not flux calibrated. We checked that the
flux in the Lyman-$\alpha$ emission line relative to the continuum
from the quasar is about the same as in the spectrum of
M{\o}ller et al. (\cite{moller}).
The measured Lyman-$\alpha$ flux from M{\o}ller et al. (\cite{moller})
is $F=5.4^{+2}_{-0.8}\times10^{-17}$ erg s$^{-1}$ cm$^{-2}$.
Assuming $H_0=70$ km s$^{-1}$ Mpc$^{-1}$, $\Omega_{\mathrm{M}}=0.3$
and $\Omega_{\mathrm{\Lambda}}=0.7$, the redshift $z=2.0400$ corresponds
to a luminosity distance
of $d_{\mathrm{L}} = 15921$ Mpc.
The observed flux therefore corresponds to a Lyman-$\alpha$ luminosity of
$L_{\mathrm{Ly\alpha}}=1.64\times 10^{42}$ erg s$^{-1}$.
Adopting the relation between the measured Lyman-$\alpha$ luminosity and the
star formation rate from Kennicutt et al. (\cite{kennicutt}),
$~L_{\mathrm{Ly\alpha}}=10^{42}\times\mathrm{SFR}$ , we derive a star formation
rate
of
SFR~=~$1.6^{+0.6}_{-0.3}~\mathrm{M_{\sun}~yr^{-1}}$.
The Lyman-$\alpha$ emission provides only a lower limit to the SFR
as the presence of dust can reduce the strength of the Lyman-$\alpha$ emission.
In the following section, we will see that the depletion factor is not small
and that dust is present in the gas.



\section{Column densities and metallicities}\label{coldens}

\begin{figure}
   \centering
   \includegraphics[bb=28 400 250
810,width=7cm,height=10cm,clip]{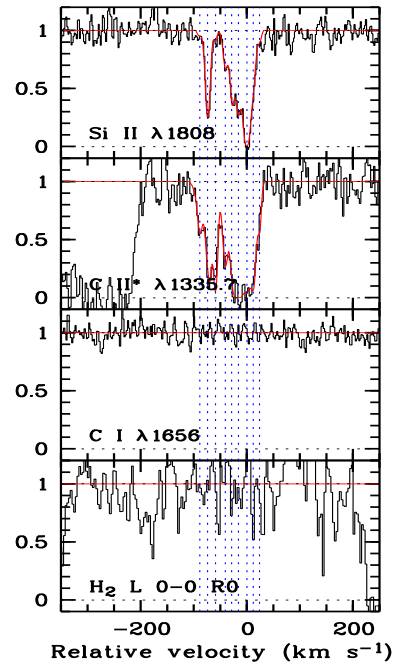}
      \caption{
               Absorption line profiles of the \ion{C}{ii$^{*}$}$\lambda$1335
               doublet and the \ion{Si}{ii}$\lambda$1808 transition.
Nine individual components were needed to perform the fit
to the absorption lines that is overplotted to the data as a solid line.
The components are indicated by vertical dashed lines. The portions of the
spectrum where the C\,{\sc i}$\lambda$1656 and H$_{2}$L0R0 transitions are
expected are also shown. The zero point of the velocity scale has
been taken at $z_{\rm abs}$~=~2.03954.}
               
         \label{fits}
   \end{figure}

\begin{figure}
   \centering
   \includegraphics[bb=55 390 400
775,width=8.5cm,clip]{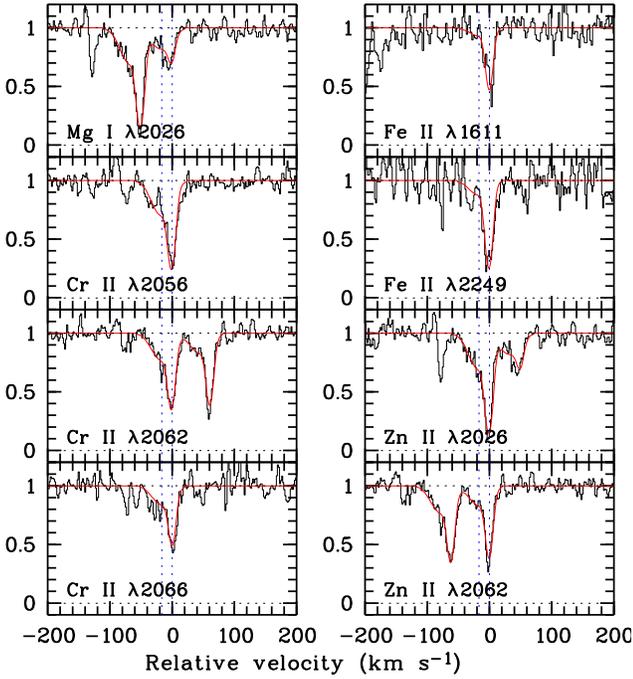}
      \caption{
               Voigt-profile fits to the \ion{Fe}{ii}, \ion{Cr}{ii},
               \ion{Zn}{ii} and \ion{Mg}{i} absorption lines.
               When lines are blended simultaneous fits were performed.
               The zero point of the velocity
scale has been taken at $z_{\rm abs}$~=~2.03954.}
         \label{fits2}
   \end{figure}

A number of metal absorption lines associated with the DLA system are
detected: low-ionization species, e.g.,
 \ion{C}{ii}$^{*}$, \ion{Si}{ii}, \ion{Fe}{ii}, \ion{Cr}{ii},
 \ion{Al}{ii} and \ion{Al}{iii}, and the high-ionization species
 \ion{C}{iv} and \ion{Si}{iv}.
 The column densities were derived via Voigt-profile fitting
of the absorption lines, using for each species the different transitions
present in the spectrum.
 Fig. \ref{fits} and Fig. \ref{fits2} show a sample of the fitted line
 profiles.
 For the strong \ion{C}{ii$^{*}$} and \ion{Si}{ii} lines, nine components
 are needed to reproduce the profiles.
Their redshifts ($z_{\rm abs}$), column densities (log $N$ in cm$^{-2}$) and
turbulent broadening parameters ($b_{\rm turb}$ in km~s$^{-1}$)
 are listed in Table \ref{table1}. The components represent gas clouds
 associated with the DLA galaxy.
 The weaker lines were only detected in the two strongest components
 at $z_{\rm abs}$~=~2.03937 and 2.03954 (labeled components 6 and 7),
 which are responsible for the main absorption
(log~$N$(\ion{Si}{ii})$>$15.2).
It is important to note that 21~cm absorptions have been reported
by Wolfe et al. (1985) in these two components. Most of the neutral hydrogen
is therefore probably associated with these two components.
The 21~cm absorption is stronger at $z_{\rm abs}$~=~2.03937 than at
$z_{\rm abs}$~=~2.03954 in accordance with the \ion{C}{ii$^{*}$}
absorption. Other species and in particular \ion{Si}{ii} show the contrary;
their column densities are higher in component 7 than in component 6.
Our column density determinations compare well with that
of Prochaska \& Wolfe (1999) except for Fe~{\sc ii} that we
find 0.2~dex less abundant based on the Fe~{\sc ii}$\lambda$1611
optically thin transition.

Assuming that these two components dominate the DLA system and contain
most of the neutral hydrogen,
 we derived the total column densities from these two
 components and the corresponding abundances relative to solar
 ($\mathrm{[X/H]=log~(X/H)-log~(X/H)_{\sun}}$). The solar values
 were taken from Morton (\cite{morton}).
 The H~{\sc i} column density
 cannot be constrained for individual components.
 Therefore, taking into account only the column density summed over
 the two strongest components of the system
can introduce a systematic error
 in the sense that our derived metallicities could be lower limits.
 We can estimate the possible corresponding error by integrating the
column densities for
 all components of \ion{Si}{ii} (Table \ref{table1})
 and comparing with the value obtained for the two strongest
 components (Table \ref{table2}).
 In this way we derive a metallicity for
 silicon of [\ion{Si}/H]~=$-$1.11 instead of $-$1.28, which means
 that the metallicity of silicon could be underestimated by at most 0.17 dex.

 From the values in Table \ref{table2}, we
 can derive abundance ratios for different metals.
In the ISM of our Galaxy, zinc and silicon are barely depleted
onto dust-grains which is consistent
with our observed ratio in this high redshift system of [\ion{Si}{ii}/\ion{Zn}{ii}]~=~$-$0.06.
Other elements appear depleted:
 chromium ([\ion{Cr}{ii}/\ion{Zn}{ii}]~=~$-$0.44),
iron ([\ion{Fe}{ii}/\ion{Zn}{ii}]~=~$-$0.65), phosphorus
([\ion{P}{ii}/\ion{Zn}{ii}]~=~$-$0.32) and manganese
([\ion{Mn}{ii}/\ion{Zn}{ii}]~=~$-$0.83).
The depletion of iron compared to zinc is indicative
of the presence of dust at a level compatible with the
presence of molecular hydrogen (see Ledoux et al. 2003).
The fact that molecular hydrogen is {\sl not} detected
(log~$N$(H$_{2}$)~$<$~14.9 and log~$f$~$<$~$-$6.52) is therefore
surprising, especially as the system has one of the highest H~{\sc i} column
densities observed in DLAs (see Section 7).

   \begin{table}
      \caption[]{Redshifts, ion column densities and turbulent
broadening parameters
                 for all components of the system}
         \label{table1}
          \centering
          \begin{tabular}{c c c c c}
          \hline\hline
          \# & $z_{\rm abs}$ & log~$N$(\ion{C}{ii$^{*}$})
         & log~$N$(\ion{Si}{ii})     & $b_{\rm turb}$\\
             &               &
         &                           & [km s$^{-1}$] \\
          \hline
  1 & 2.03864 & 13.16$\pm$0.23   & $<14.37^{\rm a}$  &
5.4$\pm$4.1\\
  2 & 2.03879 & 13.48$\pm$0.21   & 15.19$\pm$0.03  &  4.7$\pm$0.4\\
  3 & 2.03894 & 14.43$\pm$0.31 & $<14.37^{\rm a}$  &
$\sim 2.5$\\
  4 & 2.03913 & 13.55$\pm$0.15   & 14.65$\pm$0.10  &  5.2$\pm$1.8\\
  5 & 2.03925 & $<13.12^{\rm a}$   & 14.99$\pm$0.10
&  4.3$\pm$1.4\\
  6 & 2.03937 & 14.79$\pm$0.39   & 15.21$\pm$0.09  &  5.9$\pm$2.0\\
  7 & 2.03954 & 13.91$\pm$0.65   & 15.90$\pm$0.20  &  5.3$\pm$0.9\\
  8 & 2.03966 & 13.89$\pm$0.34   & 14.87$\pm$0.31  &  5.2$\pm$3.2\\
  9 & 2.03978 & $<13.12^{\rm a}$  & $<14.37^{\rm a}$
&  2.5$\pm$4.4\\
         \hline
        \end{tabular}
      \begin{list}{}{}
 \item[$^{\rm a}$] $5\sigma$ detection limit.
     \end{list}
  \end{table}

   \begin{table}
      \caption[]{Ion column densities and metal abundances in
                 components 6 and 7}
         \label{table2}
          \centering
          \begin{tabular}{c r c c}
          \hline\hline
          Species  &  log $N$(X) & $\mathrm{log(X/H)_{\sun}}+12^{\mathrm{a}}$
          & $\mathrm{[X/H]}^{\mathrm{b}}$\\
          \hline
          \ion{H}{i}        & 21.70$\pm$0.10          & ...  & ...\\
          H$_{2}$ (J=0)     & $<$14.55       & ...  & ...\\
          H$_{2}$ (J=1)     & $<$14.60       & ...  & ...\\
          \ion{C}{i}        & $<$12.45       & ...  & ...\\
          \ion{C}{ii$^{*}$} & 14.84$\pm$0.35 & ...  & ...\\
          \ion{Mg}{i}       & 13.26$\pm$0.04 & ...  & ...\\
          \ion{Mg}{ii}      & 16.07$\pm$0.07 & 7.58 &$-$1.21$\pm$0.12 \\
          \ion{Si}{ii}      & 15.98$\pm$0.17 & 7.56 &$-$1.28$\pm$0.20 \\
          \ion{P}{ii}       & 13.72$\pm$0.04 & 5.56 &$-$1.54$\pm$0.11 \\
          \ion{Cr}{ii}      & 13.73$\pm$0.02 & 5.69 &$-$1.66$\pm$0.10 \\
          \ion{Mn}{ii}      & 13.18$\pm$0.04 & 5.53 &$-$2.05$\pm$0.11 \\
          \ion{Fe}{ii}      & 15.33$\pm$0.04 & 7.50 &$-$1.87$\pm$0.11 \\
          \ion{Zn}{ii}      & 13.15$\pm$0.02 & 4.67 &$-$1.22$\pm$0.10 \\
         \hline
         \end{tabular}
     \begin{list}{}{}
     \item[$^{\mathrm{a}}$] Reference abundances from Morton (\cite{morton}).
     \item[$^{\mathrm{b}}$] The given errors correspond to errors in
         the column densities.
     \end{list}
   \end{table}


\section{Kinematics}

Figure \ref{metals} shows the velocity profiles of several low (\ion{C}{ii},
\ion{C}{ii}$^{*}$, \ion{Si}{ii}, \ion{Fe}{ii}, \ion{Al}{ii}, \ion{Al}{iii})
and high (\ion{C}{iv}, \ion{Si}{iv}) ionization metal absorption lines.
The zero point of the velocity scale is located at $z_{\rm abs}$~=~2.03954
and corresponds to the position of the strongest Fe~{\sc ii} and
Zn~{\sc ii} component (see Fig.~4).
We indicate in Fig.~5 the position
of the peak of the Lyman-$\alpha$ emission line as a vertical solid line
($\Delta v$~$\approx$~+45~km~s$^{-1}$).
As can be seen from the figure the metal absorption
lines are all blueshifted with respect to the Lyman-$\alpha$ emission.
This is a consequence of kinematics either in the disk of a
galaxy or in outflowing gas. We shall discuss both possibilities in more
detail below.

\subsection{Rotating disk}

Prochaska \& Wolfe (\cite{prochaska}) made the case that the kinematics of
DLA systems can be explained by models of large rotating disks.
For the most likely rapidly rotating thick disk model they found that
the absorption profile should be asymmetric with the strongest
absorption component located at one edge of the profile. In this model
the emission originating from the central part of the disk
should be offset from the absorptions.
This corresponds to what we observe in the DLA
towards PKS~0458$-$020. Indeed, the velocity profiles of all metal
absorption lines
are observed blueward of the Lyman-$\alpha$ emission and the
strongest absorption
component is located at the red edge of the profile. In addition,
the Lyman-$\alpha$ emission is located outside of the
absorption profile at
$\Delta v~=~45\pm16$ km s$^{-1}$ (the uncertainty comes from
the uncertainty in the redshift of the Lyman-$\alpha$ emission) redward
of the strongest absorption component.
If we assume that the line of sight goes through a large rotating disk
and the Lyman-$\alpha$ emission originates from the center of this disk,
the blueshift of the low-ionization transition lines compared to the
Lyman-$\alpha$ emission can be explained by gas that takes part in the
rotation of the disk and is moving towards us. The small velocity offset
of the
strongest absorption compared to the emission suggests that the line of
sight crosses the mid-plane
of the disk far from the major axis, where the projected
rotational velocity is small. The impact parameter between the line of sight
and the center of the disk should be small and
the inclination high to ensure
strong enough absorption spread over more than 100~km~s$^{-1}$.
A small impact parameter between the emitting region and the line of sight
was derived by M\o ller et al. (2004) of the order of 0.3~arcsec or
2.5~kpc.
It is striking that the observed situation here corresponds to
Case 4 of Figure 14 in Prochaska \& Wolfe (1997) and supports
the case for a large rotating disk. Note that the same conclusion has
been drawn from 21~cm observations by Wolfe et al. (1985).
This scenario
assumes that the Lyman-$\alpha$ emission line peak records
the systemic velocity of the galaxy. This may not be the case
as indicated by the velocity shifts usually observed between
the Lyman-$\alpha$ and the [O\,{\sc iii}] emissions in Lyman break galaxies
(Pettini et al. 2001) or in DLA systems (Weatherley et al. 2005).

\subsection{Galactic wind}

On the other hand, the observed situation is also reminiscent of a wind
flowing out
of a star-forming region in our direction as observed in star-burst
galaxies
(e.g., Veilleux et al. \cite{veill}). The velocity offsets
derived here for the strongest
absorption components seem too small to be caused by a wind.
We note that the high-ionization lines \ion{Si}{iv}
$\lambda\lambda$1393,1402
and \ion{C}{iv} $\lambda\lambda$1548,1550
show a second broad absorption component at a projected
velocity of $-$170 km s$^{-1}$. This strong feature is completely
absent in the lower ionization lines.
There are two possibilities to explain this high-ionization region.
Either the absorption comes from a region of hot gas associated with the
DLA absorber but located at the projected distance corresponding to
$v$~= $-$170 km s$^{-1}$, or there is hot gas moving towards us
with
this velocity which has been ejected by the DLA galaxy. The first
explanation is unlikely as in that case the gas should be located close
to the center of the galaxy for the rotation velocity to be large
and should therefore be associated with less ionized gas.
The observed high velocity offset component could
be a heated shock front or the galactic wind itself.
However, this is clear indication for a galactic outflow driven by the
mechanical energy deposited by supernova and stellar winds in star-forming
regions.
\par\noindent
Blueshifted absorption and redshifted Lyman-$\alpha$ emission is also seen
in high-redshift spectra of UV-selected galaxies and is also interpreted to
be caused by winds (see Adelberger et al., 2005, and references therein).
For a sample of Lyman-break galaxies, Pettini et al. (2001)
showed that the Lyman-$\alpha$ emission is redshifted
by 200 to 1100 km s$^{-1}$ relative to the position of optical emission
lines
(H$\beta$ and [O\,{\sc iii}])
and that the absorptions for three-quarters of the sample are
blueshifted by a median value of -300 km s$^{-1}$. They interpret this
as the
signature of strong galactic winds.
In this scenario, the gas seen in absorption in front of the stars is the
approaching part of an expanding shell of swept-up material that has a very
high optical depth to Lyman-$\alpha$ photons, so that the only detectable
Lyman-$\alpha$ emission along the line of sight is from the back of the
shell,
behind the stars, receding at
velocities where no foreground absorption takes place (Pettini et al.
 \cite{pettini2}). This is in agreement with the asymmetric Lyman-$\alpha$
emission line profile we observe,
although the observed velocity range of
$\Delta v$~=~120 and 200 km~s$^{-1}$ for the low and high-ionization
species respectively in the DLA towards PKS~0458$-$020 is smaller
than the above findings.
Weatherley et al. (2005) presented the detection of [\ion{O}{iii}]
emission from galaxies responsible for two other damped Lyman-$\alpha$
systems. The velocity differences between the Lyman-$\alpha$
emission and the [\ion{O}{iii}] lines is about 100 km~s$^{-1}$ in both
systems. Rest-frame optical emission lines are unaffected by resonant
scattering and provide a better measurement of the galaxy systemic velocity.
Therefore, the detection of [\ion{O}{iii}] emission lines from the
galaxy in PKS~0458$-$020 could help to pin down the systemic velocity and
to better determine the situation.

\begin{figure}
   \centering
   \includegraphics[bb=50 30 513 745,width=8.5cm,clip]{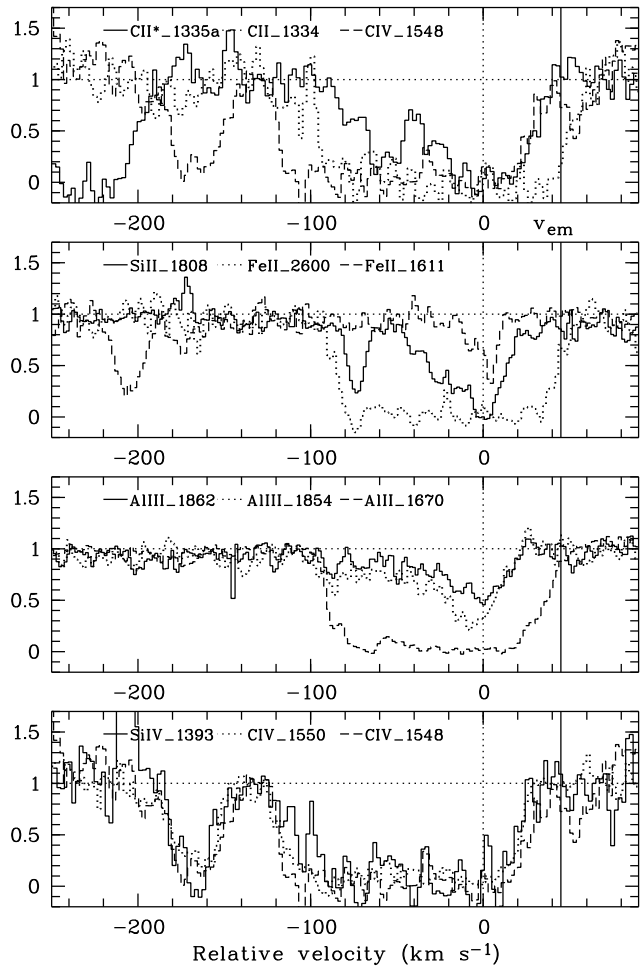}
      \caption{
                Velocity profiles of different metal absorption lines.
                The dashed vertical line marks the zero point
                derived from the Fe~{\sc ii}$\lambda$1611 absorption (see
                second panel from top). The peak of the Lyman-$\alpha$ emission
                indicated by the vertical black solid line is
                offset by about 45 km s$^{-1}$  redward
                of the main absorption component. The zero point of the
                velocity scale has been taken at $z_{\rm abs}$~=~2.03954.}
         \label{metals}
   \end{figure}

In this context, we note that the red wing of the
\ion{C}{ii}$^{*}$ $\lambda$1335
absorption profile in Fig. \ref{metals} follows exactly the red wing of
the \ion{C}{iv} $\lambda$1548 profile (the profile of the
\ion{C}{ii} $\lambda$1334 line is broader). This implies that
at least part of the \ion{C}{ii}$^{*}$ absorption comes from the warm gas
and is closely associated with the C~{\sc iv} phase. This supports
the conjecture that part of the gas is outflowing.

%

\section{The star formation rate from \ion{C}{ii}$^{*}$}

In Section \ref{lyinem}, we obtained the SFR in the DLA from the
Lyman-$\alpha$ emission line flux.
Wolfe et al. (\cite{wolfeI}) proposed another technique to derive
the SFR in
DLA systems from the strength of the \ion{C}{ii$^{*}$} absorption
and the dust-to-gas ratio under the assumption that the gas in DLAs is
heated by the same mechanism responsible for the heating of the ISM
in the Milky Way. The authors argue that under steady-state conditions
the cooling rate measured from the \ion{C}{ii}$^{*}$ absorption equals the
heating rate per H atom, which can be used to infer the SFR per unit area
$\mathrm{\dot{\psi}^{*}}$. While the \ion{C}{ii}$^{*}$ absorption
strength is measured locally along the line of sight, the derived SFR
per unit area is thought to represent the mean SFR over the whole star-forming volume in the DLA.

For PKS~0458$-$020, Wolfe et al. (\cite{wolfeI}) derived two solutions for the
SFR per unit area of $\mathrm{\dot{\psi}^{*}}\approx10^{-2}$ or
$10^{-1}~\mathrm{M_{\odot}~yr^{-1}~kpc^{-2}}$
together with a gas particle density of $\mathrm{log}~n\approx1.2$
or 0.3 cm$^{-3}$
assuming that the gas where the \ion{C}{ii$^{*}$} absorption occurs is
respectively cold or warm. Recent revision
of the UV background spectrum indicates that these
SFRs could be slightly overestimated (Wolfe 2005).
Note that, by combining the above values of the particle density
with our measured \ion{H}{i} column density, we can derive a characteristic
length scale for the light path through the \ion{H}{i} absorbing region of
$\sim$0.1 or 0.8 kpc. Obviously, the physical size of the DLA galaxy
is larger than this and therefore also the region over which the above
mean star formation rates apply.

If the observed Lyman-$\alpha$ emitting region is the only source
of heating, then we can derive the size of the heated region by equating
the SFR from the Lyman-$\alpha$ emission to that derived from
C~{\sc ii}$^*$.
This size should be of the order of $R$~=~7.2 or 2.3~kpc,
respectively for the cold or warm gas,
in order for the two SFR estimates to match. Note that this size is probably
smaller than the total size of the H~{\sc i} disk, $R_{\rm disk}$.
In the following, we try to estimate $R$. The first estimate
can be obtained by assuming that the impact parameter between
the Lyman-$\alpha$ emitting region and the line of sight corresponds
to a lower limit of $R$.
M{\o}ller et al. (\cite{moller}) estimate this impact parameter
to be $b_{\mathrm{DLA}}~=~0\farcs 3\pm0\farcs 3$.
Assuming the above cosmology (see Sect.~3), a redshift of $z=2.04$
corresponds to an angular diameter distance to the DLA of
1723 Mpc, so that the angle of $0\farcs 3$ corresponds to a
proper size of
2.5 kpc.
This value coincides with the above solution if the gas is warm.
For the cold gas solution, this rather small value can be explained by the
fact that
the measured impact parameter can lie anywhere in the
range $[0;R_{\rm disk}]$.\footnote{M{\o}ller et al. (\cite{moller})
  also give a firm upper limit
for the impact
parameter of $b_{\mathrm{DLA}}~=~0\farcs 8$, by which the object would
fall outside the
slit in their observations. This value corresponds to a proper size
of 6.7 kpc}.

Another estimate of $R$ can be derived
by using
high-resolution radio interferometry observations in front of the
extended PKS~0458$-$020 radio source. Briggs et al. (\cite{briggs})
probed several different paths through the absorbing medium and
concluded that the absorber is a disk-like structure that extends across
2$\arcsec$. This corresponds to a radius of $R_{\rm disk}$~=~8.4 kpc.
The disk should be oriented in the same direction as the radio source
and therefore in the South-Western direction whereas the Lyman-$\alpha$
emitting region is in the North-Western direction. This is consistent,
as are the kinematics, with the existence of an inclined large disk with its
center
at the location of the emitting region. Therefore a large value of $R$
is not incompatible with this model.

On this basis alone, it is therefore difficult to decide whether the gas
is cold or warm. Another way of looking at this problem
is to estimate the mean UV flux along the line of sight.


\section{Missing molecules}


Molecular hydrogen is not detected in our spectrum down to a limit
of log~$N$(H$_{2}$)~=~14.9, corresponding to a  molecular fraction
($f=\mathrm{2N(H_{2})/[2N(H_{2})+N(\ion{H}{i})]}$)
of $\mathrm{log}\,f=-6.52$. The absence of molecules is
surprising at
such a high H~{\sc i} column density (log~$N$(H~{\sc i})~=~21.7).
As the gas is probably dusty with a depletion factor of [Zn/Fe]~=~0.65,
this could be a consequence of high temperature and/or high UV background
radiation.

The temperature of the gas can be estimated from the spin temperature.
Absorption at 21 cm was reported in two components at the
same redshifts as our components 6 and 7 (Wolfe et al. \cite{wolfeIII};
see Table~1).
Combining the 21 cm absorption with the \ion{H}{i}
column density, Wolfe et al. (\cite{wolfeIII}) estimated the spin temperature
of the gas to be less than 1000 K.
Kanekar \& Chengalur (\cite{kanekar}) corrected this value
to $T_{\rm S}$~$\sim$~385$\pm$100~K. This is unusually low for a
DLA system and is in the range of Galactic values ($<$350~K; Braun \& Walterbos
\cite{braun}).

The integrated 21~cm optical depth of component 6 is about four
times larger than that of component 7 in accordance with what is
seen
  for the \ion{C}{ii$^{*}$} absorption.
It is however apparent that
the metal column densities are smaller in component 6 (see Fig.~4);
the Si~{\sc ii} column density is for example five times smaller in
component 6 than in component 7.
On the basis of this inverted ratio, if we assume similar physical
conditions (metallicity and ionization
factor) in both components, then the H~{\sc i} column density is
smaller in component 6 than in component 7 by a factor of five
and the spin temperature is smaller by a factor of twenty.
Conversely, assuming a similar temperature would lead to a metallicity
ten times smaller in component 6 which would be at odds with the
high homogeneity of the gas usually seen in DLA systems (see Rodriguez et al.
2005). The above temperature of 385~K is the harmonic mean
between the temperatures in the two components weighted by
the H~{\sc i} column densities. Given the above ratio,
we conclude that the spin temperatures of components 6 and 7 are
of the order of 2000 and 100~K respectively.

Using the SFR from the Lyman-$\alpha$ emission and the relation between
SFR and
UV flux,  ${\rm SFR}~=~L_{\rm UV}~\times~1.4\times10^{-28}$ (Kennicutt et
al. \cite{kennicutt}), we derive a specific
UV luminosity of
$L_{\rm UV}~=~1.14\times10^{28}~\mathrm{erg~s^{-1}~Hz^{-1}}$
in the frequency range between 1500 and 2800 $\rm{\AA}$. This luminosity
corresponds to a UV flux of
$F_{\rm
UV}=1.53\times10^{-17}$~erg~s$^{-1}$~cm$^{-2}$~Hz$^{-1}$
if we
assume that the distance to the absorbing gas is given by
the impact parameter.
The resulting flux is ten times larger than the UV
flux measured in the ISM of our Galaxy
($\mathrm{F_{UVgal}=1.47\times10^{-18}~erg~s^{-1}~cm^{-2}~Hz^{-1}}$
following the fit in Appendix 1 of P\'equignot \& Aldrovandi \cite{pequi}).
Therefore, the absence of molecular hydrogen is not surprising
even though
at least part of the gas is at low temperature and the depletion factor
is not
small ([Zn/Fe]~=~0.65), indicating that dust is present in the gas.
As shown by the models of Srianand et al. (2005), the absence of H$_2$ can be
explained as the consequence of the high radiation field.

Note that this is consistent with the high mean ambient UV flux derived by
Wolfe et al. (2004; $>$19.1 times the Galactic value).


\section{Summary and conclusions}

We have presented the high-resolution spectrum of the strong
damped Lyman-$\alpha$ absorber at $z_{\rm abs}$~=~2.03954 in front of
PKS~0458$-$020.
It is one of the rare systems where Lyman-$\alpha$ is clearly seen in
emission with an impact parameter between the emitting region
and the line of sight of $\sim$0.3~arcsec (M\o ller et al. 2004)
or 2.5~kpc for the adopted concordance cosmology.
We determined the redshift of the Lyman-$\alpha$ emission line to be $z=2.0400
\pm 0.0002$. The metal absorption lines are found to be blueshifted
compared to the Lyman-$\alpha$ emission and to span a velocity
range of $\Delta v$~=~120
and 200 km~s$^{-1}$ for the low and high-ionization species respectively.
The kinematics, together with the observations in 21~cm by Briggs et
al. (1989), are strikingly consistent with the model
of a large rotating disk presented by Prochaska \& Wolfe (1997)
in which the line of sight crosses the mid-plane far from the
center of the disk but keeps a low impact parameter with the center of
the disk.
Conversely, the coincidence of the red wings of the C~{\sc iv}
and C~{\sc ii}$^*$ profiles and the extent of the C~{\sc iv}
absorption argues for the presence of blueshifted warm gas
possibly part of an outflow from the DLA galaxy. If true, the
velocity of such outflows is much smaller than what is observed
in star-burst galaxies (see Veilleux et al. 2005).
A possible detection of rest-frame optical emission lines could
help
to support the model of a large rotating disk.

We derived column densities and metallicities for a number of
species. The DLA absorber corresponds to a two-phase medium with warm
and cold
gas.
We could compare the star formation rate derived from the Lyman-$\alpha$
emission line with the derivation from the \ion{C}{ii}$^{*}$ method in
the same object.
 From the Lyman-$\alpha$ emission, we find a star formation rate of
SFR~=~$1.6^{+0.6}_{-0.3}~\mathrm{M_{\sun}~yr^{-1}}$.
 From the C~{\sc ii}$^*$
column density, Wolfe et al. (2003) derived a
SFR per unit area of 10$^{-2}$ and 10$^{-1}~\mathrm{M_{\odot}~yr^{-1}~kpc^{-2}}$ for
respectively cold and warm gas. This means that the diffuse gas should be extended
over a radius of $\sim$7.2 or 2.3~kpc respectively for both
SFRs to match.

The absence of molecular hydrogen to a limit of
log~$f$~=~$-$6.52  can be explained as the consequence of the high
radiation field in the disk due to star formation. The ambient UV flux due
to the observed emitting region is one order of magnitude larger than
the flux in our Galaxy.

\begin{acknowledgements}
JH and SC acknowledge the support of the European Community under a
Marie Curie Host Fellowship for Early Stage Researchers Training under
contract number MEST-CT-2004-504604.
RS and PPJ gratefully acknowledge the Indo-French Centre
for the Promotion of Advanced Research
(Centre Franco-Indien pour la Promotion de la Recherche Avanc\'ee)
under contract No. 3004-3.
PPJ thanks IUCAA for hospitality during the time
part of this work has been completed.
We thank the referee, Steve Warren, for a careful reading
of the manuscript and for providing comments that helped to improve the
paper.
\end{acknowledgements}



\begin{thebibliography}{}

  \bibitem[2005]{adel} Adelberger, K. L., Shapley, A. E., Steidel, C. C.,
       Pettini, M., Erb, D. K., \& Reddy, N. A. 2005, ApJ, 629, 636

  \bibitem[2000]{ballester} Ballester, P., Dorigo, D., Disar{\` o},
        A., Pizarro de La Iglesia, J.~A., Modigliani, A., \& Boitquin, O.
        2000, ASPC, 216, 461

  \bibitem[1992]{braun} Braun, R., Walterbos, R. A. M. 1992,
       ApJ, 386, 120

  \bibitem[1989]{briggs} Briggs, F. H., Wolfe, A. M., Liszt, H. S., et
al. 1989,
       ApJ, 341, 650

  \bibitem[2003]{chen} Chen, H.-W.,\& Lanzetta, K. M. 2003, ApJ, 597, 706

  \bibitem[2002]{curran} Curran, S. J., Webb, J. K., Murphy, M. T.,
Bandiera, R.,
       Corbelli, E., \& Flambaum, V. V. 2002,
       PASA, 19, 455

  \bibitem[2000]{dekker} Dekker, H., D'Odorico, S., Kaufer, A., Delabre, B.,
        \& Kotzlowski, H. 2000,
        SPIE, 4008, 534

  \bibitem[2003]{kanekar} Kanekar, N., \& Chengalur, J.N. 2003,
       A\&A, 399, 857

  \bibitem[1998]{kennicutt} Kennicutt, R.C. 1998,
       ARA\&A, 36, 189

  \bibitem[1997]{lebrun} Le Brun, V., Bergeron, J., Boiss\'e, P.,
  Deharveng, J. M. 1997,
       A\&A, 321, 733

   \bibitem[2002]{ledoux} Ledoux, C., Bergeron, J., \& Petitjean, P. 2002,
       A\&A, 385, 802

   \bibitem[2003]{ledoux2} Ledoux, C., Petitjean, P., \& Srianand, R. 2003,
       MNRAS, 346, 209

  \bibitem[2004]{moller} M{\o}ller, P., Fynbo, J.P.U.,\& Fall, S.M. 2004,
       A\&A, 422, L33

  \bibitem[1998]{moller2} M{\o}ller, P., Warren, S. J., Fynbo, J. P 1998,
       A\&A, 330, 19

  \bibitem[1993]{molwar} M{\o}ller, P., \& Warren, S. J. 1993,
       A\&A, 270, 43

  \bibitem[2003]{morton} Morton, D. C. 2003
       ApJS, 149, 205

  \bibitem[1986]{pequi} P\'equignot, D., \& Aldrovandi, S. M. V. 1986,
       A\&A, 161, 169

  \bibitem[2000]{petitjean} Petitjean, P., Srianand, R.,\& Ledoux, C. 2000
       A\&A, 364, L26

  \bibitem[2001]{pettini2} Pettini, M., Shapley, A. E., Steidel, C. C.,
       Cuby, J.G., et al. 2001
       ApJ, 554, 981

  \bibitem[1994]{pettini} Pettini, M., Smith, L. J., Hunstead, R. W., \&
       King, D. L. 1994
       ApJ, 426, 79

  \bibitem[1997]{prochaska} Prochaska, J.X.,\& Wolfe, A.M. 1997
       ApJ, 487, 73

 \bibitem[1999]{prochaskaII} Prochaska, J.X.,\& Wolfe, A.M. 1999
       ApJS, 121, 369

  \bibitem[2005]{rodriguez} Rodriguez, E., Petitjean, P., Aracil, B., Ledoux, C., \&
Srianand, R.
2005, A\&A, accepted

 \bibitem[2005]{Srianand} Srianand, R., Shaw, G., Ferland, G.,
Petitjean, P.,
\& Ledoux, C. 2005, astro-ph/0506556

  \bibitem[2005]{veill} Veilleux, S., Cecil, G., \& Bland-Hawthorn, J. 2005
       Astro-ph/0504435

  \bibitem[2004]{vreeswijk} Vreeswijk, P.M. et al. 2004,
       A\&A, 419, 927

  \bibitem[2001]{warren} Warren, S. J.,  M{\o}ller, P., Fall, S. M., \&
       Jakobsen, P. 2001
       MNRAS, 326, 759

 \bibitem[2005]{weather} Weatherley, S. J., Warren, S. J., Møller, P.,
       Fall, S. M., Fynbo, J. U.,\& Croom, S. M. 2005,
       MNRAS, 358, 985

 \bibitem[2005]{wolfe05} Wolfe, A. M., 2005, in ``Probing Galaxies
Through Quasar
Absorption Lines''. IAU Symposium 199, eds P. R. Williams, C, Shu, B.
M\'enard, Shanghai

\bibitem[1985]{wolfeIII} Wolfe, A. M., Briggs, F. H., Turnshek, D. A.,
       Davis, M. M., Smith, H. E., \& Cohen, R. D. 1985,
       ApJ, 294, L67

 \bibitem[2004]{wolfe04} Wolfe, A.M., Howk, J. C., Gawiser, E.,
Prochaska, J.X.,\&
Lopez, S. 2004, ApJ 614, 625

\bibitem[2003]{wolfeI} Wolfe, A. M., Prochaska, J. X., \& Gawiser, E. 2003,
       ApJ, 593, 215




\end{thebibliography}
\end{document}